\documentclass[preprint,preprintnumbers,amsmath,amssymb,nofootinbib]{revtex4}

\usepackage{amssymb,amsthm,amscd,amsbsy,array}
\usepackage{bm}
\usepackage{soul} 
\usepackage{graphics,graphicx,xcolor}
\usepackage[colorlinks=true, pdfstartview=FitV, linkcolor=blue, citecolor=blue, urlcolor=blue]{hyperref}

\usepackage{amsfonts}
\usepackage{latexsym}
\usepackage{etex}
\usepackage{dcolumn}
\usepackage{amsmath,dsfont}
\usepackage{soul}

\def\ben{\begin{equation}}
\def\een{\end{equation}}
\def\bea{\begin{eqnarray}}
\def\eea{\end{eqnarray}}

\def \babla{\boldsymbol{\nabla}}

\newcommand{\bx}{{\bm{x}}}

\renewcommand{\Re}{\mathrm{Re}}

\newcommand{\bv}{{\bf v}}

\def\bv{{\bm{v}}}

\def\beq{\begin{equation}}
\def\eeq{\end{equation}}
\def\beqa{\begin{eqnarray}}
\def\eeqa{\end{eqnarray}}

\def\barray{\left(\begin{array}}
\def\earray{\end{array}\right)}
\def\barraynb{\begin{array}}
\def\earraynb{\end{array}}

\def\besub{\begin{subequations}}
\def\balign{\begin{align}}
\def\esub{\end{subequations}}
\def\ealign{\end{align}}

\def\?{\quad{\gb{\fbox{\texttt{?}}\;}}\quad}
\def\p{{\partial}}

\def\v0{\mathbf{0}}

\usepackage{color}

\newcommand{\gb}{\colorbox{green}}

\newcommand{\bblue}{\begin{bluetext}}
\newcommand{\eblue}{\end{bluetext}}
\newcommand{\bred}{\begin{redtext}}
\newcommand{\ered}{\end{redtext}}

\def\beq{\begin{equation}}
\def\eeq{\end{equation}}
\def\bea{\begin{eqnarray}}
\def\eea{\end{eqnarray}}

\def\p{\partial}

\def \p{{\partial}}

\def\semidirectproduct{
{\ooalign
{\hfil\raise.07ex\hbox{s}\hfil\crcr\mathhexbox20D}}} 

\def\6{\partial}
\def\7{\tilde}
\def\8{\widehat}

\def\G11{\Gamma_{11} }

\usepackage{color}

\newenvironment{redtext}{\color{red}}{\ignorespacesafterend}
\newenvironment{bluetext}{\color{blue}}{\ignorespacesafterend}

\def\beq{\begin{equation}}
\def\eeq{\end{equation}}
\def\beqa{\begin{eqnarray}}
\def\eeqa{\end{eqnarray}}

\def\and{{\quad\text{and}\quad}}

\newcommand{\const}{\mathop{\rm const.}\nolimits}
\newcommand{\half }{\frac{1}{2}}

\newcommand{\mybox}[1]{\fbox{$\;\displaystyle{#1}\;$}}

\let\ssection=\section
\renewcommand{\section}{\setcounter{equation}{0}\ssection}

\numberwithin{equation}{section}

\let\ssection=\section
\renewcommand{\section}{\setcounter{equation}{0}\ssection}

\begin{document}

\preprint{arXiv:1802.03370v3 [gr-qc]}

\title{Cosmological aspects of the Eisenhart-Duval lift
\\[6pt]
}

\author{
M. Cariglia$^{1}$\footnote
{e-mail: marco@iceb.ufop.br},
A. Galajinsky$^{2}$\footnote{e-mail: galajin@tpu.ru},
G.W. Gibbons$^{3}$\footnote{
e-mail: G.W.Gibbons@damtp.cam.ac.uk},
P.A. Horvathy$^{4}$\footnote{e-mail: horvathy@lmpt.univ-tours.fr}
}
\affiliation{
$^{1}${\small DEFIS, Universidade Federal de Ouro Preto,  MG-Brasil}
\\
$^2${\small Tomsk Polytechnic University, Russia}
\\
$^3${\small D.A.M.T.P., Cambridge University, U.K.}
\\
$^4${\small Laboratoire de Math\'ematiques et de Physique Th\'eorique, Universit\'e de Tours, France}
\\
}

\date{\today}

\begin{abstract}
A cosmological extension of the Eisenhart-Duval metric is constructed by incorporating a cosmic scale factor and the energy-momentum tensor into the scheme. The dynamics of the spacetime is governed the Ermakov-Milne-Pinney equation.
Killing isometries include spatial translations and rotations, Newton--Hooke boosts and translation in the null direction. Geodesic motion in Ermakov-Milne-Pinney cosmoi is analyzed.
The derivation of the Ermakov-Lewis invariant, the Friedmann equations and the Dmitriev-Zel'dovich equations within the  Eisenhart--Duval framework is presented.
\\[40pt]
\\[40pt]
\noindent\textbf{Keywords}:
Cosmology, Eisenhart-Duval lift, Ermakov-Milne-Pinney equation, time-dependent oscillator, Newton-Hooke group.
\\
\end{abstract}

\maketitle

\baselineskip=16pt

\tableofcontents

\newpage
\section{Introduction}

The Eisenhart-Duval lift
\cite{Eisenhart,Bargmann} provides a convenient framework for treating time dependent
dynamical systems and their symmetries \cite{Cariglia:2016oft}. Such systems include time-dependent harmonic oscillators.
time-dependent harmonic oscillators and their symmetries have a large literature. In particular, they are conveniently studied
using the Ermakov-Milne-Pinney  equation \cite{Ermakov,Milne,Pinney}.

An important tool for solving
the Ermakov-Milne-Pinney equation is a constant of the motion
known as the \emph{Ermakov-Lewis invariant} -- which is in fact an exact invariant for any time-dependent harmonic oscillator.
It is then natural to inquire about
 the relationship between the Eisenhart-Duval lift and the
Ermakov-Milne-Pinney equation. Can one obtain the Ermakov-Lewis invariant using the Eisenhart-Duval lift?

There is also a considerable body of
work in which methods based on the Ermakov-Milne-Pinney equation have been
applied to cosmology \cite{Rosu:1998wr,Rosu:1999hp,Rosu:1999ud,Rosu:1999rs,Hawkins:2001zx,Williams:2003vb,Rosu-Espinoza,Graham:2011nb}, in particular  to
finding solutions and the symmetries of the Friedmann equations.
This leads to the further question of what
 insights can be gained into cosmology using the Eisenhart-Duval lift?

A first attempt to incorporate a cosmic scale factor into the Eisenhart-Duval scheme has been reported  recently  \cite{Galajinsky}. Yet, since these considerations relied upon the Ermakov-Milne-Pinney equation with a constant frequency, the resulting spacetime turned out to be stationary. It is then natural to wonder whether a \emph{dynamical} spacetime with cosmological features can also be constructed along the same lines.

The purpose of this paper is to address the
questions raised above, focusing on the cosmological aspects of the Eisenhart-Duval lift.

The paper is organized as follows.
In section \ref{EMPeq}  we provide a thorough review of the Ermakov-Milne-Pinney equation for readers who may not be familiar with the rather extensive literature on this subject.

In subsection \ref{ELCMS}  a brief account of the Eisenhart-Duval lift  is given. It is shown that any Newtonian mechanical system can be represented by a metric which solves the Einstein equations provided the energy momentum tensor is chosen in a suitable way.
In subsection \ref{LewisSect}
 the Ermakov-Lewis constant is obtained by using a conformal transformation of the associated Eisenhart-Duval metric.
An explicit formula which depends upon a solution of the Ermakov-Milne-Pinney equation is presented.

Then in subsection \ref{EMPEDliftsec}  the Eisenhart-Duval lift involving a scale factor analogous to that in the Friedmann metrics is given.
It is shown that the dynamics of the spacetime is governed by the Ermakov-Milne-Pinney equation.
Subsection \ref{ConfKillingSec} contains a detailed derivation of the conformal symmetries
of this modified metric. Geodesic motion in  Ermakov-Milne-Pinney cosmoi is analyzed in subsection \ref{GeoMot}.

Subsection \ref{MattMod} contains some introductory material on Friedmann-Lema\^\i tre-Robertson-Walker (FLRW) metrics, establishing the terminology and the conventions. The Einstein equations reduce to
the Friedmann and Raychaudhuri equations; in the case of radiation and dark energy, they reduce to the Ermakov-Milne-Pinney equation.

Subsection \ref{CoClocks} deals with conformal symmetries and what we refer to as temporal diffeomorphisms, which may be used to provide explicit solutions to the Friedmann equations. These  cases include the reduction to a time-independent harmonic oscillator.

Subsection \ref{EisenFLRWSec} contains a derivation of the Eisenhart-Duval lift of geodesic motion in FLRW spacetimes. This allows describing the system's dynamical symmetries, and relating them to the null geodesics of the Ermakov-Milne-Pinney spacetime of subsection \ref{GeoMot}. The result is generalized to obtain the lift of the Dmitriev-Zel'dovich equations.

Finally in subsection \ref{EisenFriedSec} we regard the Friedmann equations as a dynamical system with a constraint and provide a generalised Eisenhart-Duval lift of it.

In section \ref{concl} we summarize our results.

Throughout the paper summation over repeated indices is understood.

\section{The Ermakov-Milne-Pinney Equation}\label{EMPeq}

In this section we shall give a brief account of
the Ermakov-Milne-Pinney equation \cite{Ermakov,Milne,Pinney} and of its uses.
The most important of these are:
\begin{itemize}
\item In dynamics: using its solutions to
map problems involving  time-dependent harmonic oscillators to problems involving time-independent harmonic oscillators.
\item In quantum mechanics; using its solutions to
 map   problems involving the one dimensional time independent Schr\"odinger
equation into Schr\"odinger
equations which have explicit solutions.
\end{itemize}
For historical remarks about the origins
of the subject the reader may consult \cite{MorrisLeach}.
More ambitious in scope is the Master Thesis \cite{Padilla}.
The reader should be aware that in many papers either Milne, or Pinney is omitted from the name of the equation.  The order of authors used here merely reflects the order of publication.

\subsection{Comparison of two time-dependent harmonic oscillators}

This is  perhaps the most direct approach to the EPM equation. Apparently due originally to Dingle \cite{Dingle}, this proceeds as follows \cite{Kamenshchik:2005kf}. We have two time-dependent harmonic oscillator equations
\besub
\begin{align}
\frac{d^2q}{dt^2 } + \omega^2(t) q&=0
\label{I} \,,
\\
\frac{d^2Q}{d\tau ^2} + \Omega^2(\tau) Q &=0 \label{II}\,.
\end{align}
\label{I-II}
\esub
which may be obtained by varying respectively  the two integrals
\begin{subequations}
\begin{align}
&\int \half \Bigl( \big (\frac{dq}{dt} \big ) ^2 - \omega^2(t) q^2\Bigr )  dt \,,
\\[6pt]
& \int \half \Bigl( \bigl(\frac{dQ}{d\tau} \bigr ) ^2  -
\Omega^2(\tau) Q^2 \bigr) d \tau \,.
\end{align}
\label{lagrangians}
\end{subequations}
We seek a  temporal diffeomorphism $\tau = \tau(t)$
taking one integral into the other up to a boundary term. If
\ben
f(t)= \bigl(\frac{d \tau}{dt }\bigr )^{-\half} \,, \label{A}
\een
and
\ben
q(t) = \bigl(\frac{d\tau}{dt}\bigr )^{-\half} Q(\tau) = f(t) Q(\tau) \label{B}
\een
we find that
\ben
\int \Bigl( \bigl( \frac{dQ}{d\tau}\bigr ) ^2 -\Omega^2(\tau) Q^2\Bigr )  d \tau =
\int \half \Bigl(  \bigl (\frac{dq}{dt} \bigr )^2  - \omega^2(t) q^2 \Bigr )  d t
-\int  \frac{d}{dt} \Bigl ( q^2 \frac{1}{f^2} \frac{df}{dt}  \Bigr ) dt,
\een
where
\ben
\omega^2(t)  = \frac{\Omega ^2(\tau)}{f^4 }    -
\frac{1}{f}\frac{d^2f}{dt^2}
\label{Dingle1}
\een
or
\ben
\frac{d^2f}{dt^2} + \omega ^2 f = \frac{\Omega ^2(\tau)}{f^3} \,.
\label{Dingle2}
\een

This equation may be read in different ways.
\begin{itemize}
\item If the solutions  $Q(\tau)$  of  equation
(\ref{II}) are known and we want
  the solutions $q(t)$  of  equation (\ref{I}),
we need to  solve  for $f(t)$ and hence
obtain $\tau(t)$ by integration
\ben
\tau = \int \frac{dt}{f^2}\,.
\label{diff}
\een

\item Conversely, if the solutions $q(t)$ of  (\ref{I}) are known and we want the solutions  of equation (\ref{II})
we need to solve equation (\ref{Dingle2}) for $f(t)$.
\end{itemize}

In both cases the procedure is complicated by the need to express $\Omega ^2 (\tau)$ as a function of $t$. Since this involves
the integral (\ref{diff}) giving $\tau(t)$, this leads in general to very complicated equations, for which the only feasible strategy appears to
 involve extensive iterations of the type described in
\cite{Kamenshchik:2005kf}\,.

\subsection{The Ermakov-Milne-Pinney equation and the Ermakov-Lewis invariant}\label{EMP-L}

Considerable simplification results if we assume
\ben
\Omega^2(\tau)= \lambda,
\een
where $\lambda$ is a constant which need not   necessarily be positive.
This leads to the original form of the \emph{Ermakov-Milne-Pinney equation:}
\ben\mybox{
\frac{d^2f}{dt^2} + \omega^2(t)f = \frac{\lambda}{f^3}
\,,
}\label{EMP}
\een
which may be thought of as a time-dependent harmonic oscillator with a non-linear term.
We remark here that  the inverse-square potential is known to have the Schr\"odinger-type 2:1 scaling symmetry \cite{DFF}.

The Ermakov-Milne-Pinney equation has the remarkable property \cite{Ermakov,Pinney}
that its general solution  may
be expressed in terms of two linearly independent solutions,
$q_1(t)$ and $q_2(t)$, of
(\ref{I}).
\ben\label{PRO1}
f(t)= \Bigl(A q_1^2(t) + B q^2_2(t) + 2C q_1(t)q_2(t)  \Bigr)^\half
\een
where $A,B,C$ are constants satisfying
\ben\label{PRO2}
AB-C^2 = \frac{\lambda}{W^2} \,,
\qquad
W= q_1 \frac{dq_2}{dt} - q_2 \frac{dq_1}{dt}\,.
\een
$W$ is the constant Wronskian of the pair of linearly independent solutions $q_1\,,q_2$.


Given a solution of the Ermakov-Milne-Pinney equation (\ref{Dingle2})
one has the identity
\besub
\begin{align}
H&=\half \Bigl( (\frac{dQ}{d\tau})^2 - \lambda Q^2 \Bigr)
\label{Hamilton}
\\[6pt]
&= \half \Bigl( \frac{q^2}{f^2}+(\frac{dq}{dt}f -\frac{df}{dt} q)^2 \Bigr ) =I
\label{Lewis}\,,
\end{align}
\label{HamLewis}
\esub
which \emph{equates the conserved Hamiltonian (\ref{Hamilton}) of the time-independent harmonic oscillator (\ref{II})
with the conserved  {\it Ermakov-Lewis invariant}  (\ref{Lewis})
of the  time-dependent harmonic oscillator (\ref{I})} \cite{Lewis1,Lewis2}.
This derivation is essentially  the same as the one given in \cite{Padmanabhan:2017rsj}.

One may regard the Lewis invariant as an exact form of
the adiabatic invariant for a  time-dependent harmonic oscillator, and as such, has obvious applications to the quantum theory as will be seen shortly.

Subsequent to the paper by Lewis, the idea  was extended to charged particles moving in a spatially uniform magnetic field and a time dependent harmonic potential \cite{Lewis3,Lewis-Riesenfeld}.
This is in effect covered by the  theorems of Larmor and Kohn which allow
one to eliminate the magnetic field by passing to
a rotating frame \cite{Gibbons:2010fb,Zhang:2011bm}. Despite its great
experimental, astronomical and technological importance, we shall not pursue this issue further.

We mention for completeness that generalizations of the Ermakov-Lewis invariant (\ref{Lewis}) were
 obtained  by considering  modifications
of the pair of Ermakov-Milne-Pinney equations (\ref{I}) and (\ref{EMP})
\cite{RayReid1, RayReidLutsky}.

Further insight is provided by the observation by Eliezer and Gray \cite{EliezerGray} that
two time-dependent harmonic oscillators of the form (\ref{I}), $q_i(t)$ and $q_2(t)$,
may be considered as a particle in the  Euclidean plane $\mathbb{E}^2$,  moving under the influence of an axisymmetric harmonic potential. Moreover, if one introduces polar coordinates $r,\theta$ such that $q_1+ i q_2=r\,e^{i\theta}$, the equations of motion of
the Lagrangian
\ben
L= \half \bigl(\dot r ^2  + r^2 \dot \theta ^2 - \omega^2 (t) r^2 \bigr),
\een
imply the conservation of the angular momentum (per unit mass)
\ben r^2 \dot \theta =h \,.
\label{angmom}
\een
Then substitution back to the radial equation of motion gives
\ben
\ddot r + \omega ^2(t) r = \frac{h^2}{r^3} \label{pseudEMP}
\een
which, in the case $\lambda = h^2 >0$, coincides with the Ermakov-Milne-Pinney equation (\ref{EMP}) if $r=f$.

If one substitutes
$
q=r \cos \theta\,,\, f= r \,, \,  \lambda = h^2
$
into (\ref{Lewis}), one finds that the Ermakov-Lewis quantity is in fact the square of the angular momentum\footnote{
The case $\lambda<0$ is obtained by replacing the
Euclidean plane $\mathbb{E}^2 $ by the Minkowski plane  $\mathbb{E}^{1,1}$ \cite{Barrett:1993yn}.}
\ben
I= h^2 \,.
\een

A list of cases for which there are explicit solutions of the radial equation (\ref{pseudEMP}) are given in \cite{EliezerGray,ZP}.
Because of its application to cosmology which we will describe later, we reproduce it here.

\begin{itemize}
\item
\ben
\omega={\rm constant} =\omega_0 \,, \qquad  r^2= (A \cos \omega_0t + B \sin  \omega _0 t)^2
+ \frac{h^2}{A^2 \omega_0^2} \sin^2 \omega_0 t \,.
\een
The orbit in the plane is an ellipse.

\item
\ben
\omega = \frac{b}{t^2} \,, \qquad \theta - \theta_0= -\frac{b}{t}
\,, \qquad r= \sqrt{\frac{h}{b}}\, t
=\frac{ \sqrt{ hb}}{\theta_0 -\theta} \,.
\een
\item
\ben
\omega = \frac{b}{t} \,,\quad r^2 = bt \,,\quad b=\frac{h}{\sqrt{b^2-\frac{1}{4} }}\,,\quad  b  \ne \half \,, \quad \theta =\theta_0 + \frac{h}{b}\ln t \,,\quad
r= b \exp{\frac{b}{h} (\theta-\theta_0) } \,,
 \een
which yields a logarithmic spiral.

\item
  \ben
  \omega= bt^k \,,\qquad r^2 =\pi h A t \Bigl( J^2_n( 2 A b^2 t^{k+1} ) +
  Y^2_n( 2 A b^2 t^{k+1}) \Bigr) \,,\quad n= \half (k+1) \,,\quad
 \een
  where $J_n\,, Y_n$ are Bessel functions.
 If $k>0$ , then $r \rightarrow
 0 $ as $ t \rightarrow \infty$, and if $k<0$  then $r \rightarrow \infty $ as $ t \rightarrow \infty$.
\end{itemize}

\subsection{The one-dimensional Schr\"odinger equation and the
Liouville-Green-Jeffreys-Wentzel-Kramers-Brillouin approximation}

If we now think of the temporal  coordinates $t,\tau$ as spatial
coordinates $x,X$, the complexified   positions $q(t),Q(\tau)$ as wave functions
$\psi(x),\Psi(X)$ and the frequencies $\omega(x)$ and $\Omega(\tau)$ as wave vectors $k(x)$ and $K(x)$, we  may transcribe  all of the fore-going theory into quantum mechanical language. The time-dependent harmonic oscillator equations (\ref{I})-(\ref{II}) become
\besub
\begin{align}
&\frac{d^2 \psi(x)}{dx^2} + k^2(x) \psi =0 \,,
\label{IQ}
\\[6pt]
&\frac{d^2 \Psi(x)}{dX^2} + K^2(X) \Psi =0 \,,
\label{IIQ}
\end{align}
\esub
where $k^2 = 2 (E-V(x)) $ and we have used units in which $\hbar^2 = m$,
$E$ is the energy, and $V(x)$ the potential energy.

The relation to the Liouville-Green-Jeffreys-Wentzel-Kramers-Brillouin method may be seen by
noting  that (\ref{A}) and (\ref{B}), (\ref{diff}),
and (\ref{Dingle2}), are identical to (2), (6), (3) in section {17.122} of \cite{Jeffreys},
 which  contains  a systematic procedure for obtaining asymptotic expansions, provides a list of references and justifies the addition of the names of Liouville, Green and Jeffreys
to the better known Wentzel, Brillouin, and Kramers.

If we choose $K^2 =X^2$ we find formal  bound state solutions of (\ref{IQ}) which take the form
\ben
\psi(x) \propto r(x) \sin \Bigl(\int^x_{x_o}\frac{d \tilde x}{r^2(\tilde x)}
\Bigr) \,,
\label{boundstates}
\een
where the dependence of $r(x)$ on $E$ has been suppressed.
Eqn (\ref{boundstates}) implies the {\it Milne quantization condition}\footnote{
This entitles Milne to his place in the name of (\ref{EMP}) \cite{Ioffe:2002tk}.}
for the allowed values of the energy $E$ \cite{Milne},
\ben
\frac{1}{\pi} \int_{-\infty}^{\infty}
\frac{dx}{r^2(x,E)} = n+1,\,\qquad n=0,1,2,3,\dots \,,
\label{MQC}
\een
where we have re-instated the dependence of $r$ on $E$ as well as $x$ \footnote{Note that $\frac{\hbar^2}{2m}$
in (3) of \cite{Ioffe:2002tk}  should be replaced by its inverse.}.

The quantum mechanical treatment presented above
was based on the formal equivalence of the time-dependent harmonic oscillator equation (\ref{I})
and the time-independent Schr\"odinger equation (\ref{IQ});
effected by exchanging  the time coordinate for the spatial coordinate
of the latter.  A deeper connection may be obtained by considering
a wave packet solution of the time dependent
Schr\"odinger equation in one spatial dimension
\cite{Schuch:2008tha,Guerrero:2013bva,Lopez-Ruiz:2014vpa,Kim:2016iem},
\ben
i \hbar \frac{\p \Psi(x,t)}{\p t}=
\Big[-\frac{\hbar ^2}{2m} \frac{\p ^2}{\p x ^2} + \half m \omega^2(t) x^2 \Big]\,\Psi(x,t)
\,
\label{TDSeqn}
\een
for
\ben
\Psi(x,t)= N(t) \exp\big[i \bigl(y(t) \tilde x^2 +  K(t)
\bigl)\big] \,,
\label{GWP}
\een
where $\tilde x= x-q(t)$, and
\ben
q(t)  \langle x \rangle = \int^\infty_{-\infty}  \bar \Psi x \Psi dx
\een
is the  real valued expectation value of  the position of the ``classical''
trajectory and $\langle p \rangle = m \dot q $ is
the classical momentum;
the functions $N(t)$ and $K(t)$ not relevant for the following discussion.
In \cite{Schuch:2008tha} it is claimed that substitution
of (\ref{GWP}) into (\ref{TDSeqn}) leads to the time-dependent harmonic oscillator equation (\ref{I})
and the complex Riccati-type equation
\ben
\frac{2 \hbar }{m} \dot y + (\frac{2 \hbar}{m}y ) ^2 + \omega ^2 t=0 \,.
\label{CReqn}
\een
Moreover, if
\ben
\Re \,y= \frac{m}{2 \hbar} = \frac{1}{f^2},
\een
then $f$ satisfies the Ermakov-Milne-Pinney equation (\ref{EMP});
the Ermakov-Lewis quantity $I$ given by   (\ref{Lewis})
is a dynamical invariant.

We remark \textit{en passant} that the Ermakov-Milne-Pinney equation may be incorporated into the formalism
of supersymmetric quantum mechanics  \cite{Schuch:2008tha,Ioffe:2002tk}.

\subsection{The Ermakov-Milne-Pinney equation and Madelung's  hydrodynamic transcription}
\label{Madelung}

The time-dependent  Schr\"odinger equation with a general potential $V(\bx)$,
\ben
i \hbar\frac{\p \Psi}{\p t} = -\frac{\hbar^2}{2m} \babla ^2 \Psi + V \Psi \,,
\een
can be brought into the hydrodynamical form~:
Setting
$
\Psi = \sqrt{\rho}\, e^{i\frac{S}{\hbar}} \,,
$
defining the velocity
$
\bv = \frac{d\bx}{dt} = \frac{1}{m} \babla S
$
and taking real and imaginary parts,
one gets equations which resemble Euler's equations for a compressible irrotational fluid,
called a \emph{Madelung fluid} \cite{Madelung}
\besub
\begin{align}
\frac{\p \rho}{\p t}+ \babla \cdot (\rho \bv) &=0
\\
\frac{\p \bv}{\p t}+ \bv \cdot \babla \bv & = - \frac{1}{m} \babla \tilde V.
\end{align}
\label{Madel}
\esub
For a stationary state one has
$
\frac{\p S}{\p t}=-E \,, \; \frac{\p \rho}{\p t}=0\,.
$

One may also require  that the probability
density be independent of time \cite{Schuch:2008tha}. In general one typically assumes that $\babla S =0$
and hence $\sqrt{\rho}$ satisfies the  \emph{linear}  equation
\ben
- \frac{\hbar ^2}{2m} \babla ^2 \sqrt{\rho} + V \sqrt{\rho} =E \sqrt{\rho}\,.
\een
 On the other hand in one spatial dimension,
 taking note of (\ref{con}),  one may alternatively  assume
\ben
\frac{dS}{dx}= \frac{C}{\rho}
\een
for some constant $C$.
This leads to the \emph{non-linear} Ermakov-Milne-Pinney equation (\ref{EMP})  for $ f= \sqrt{\rho(x)}$  \cite{Schuch:2008tha},
\ben
\frac{  d^2 \sqrt{\rho} }{dx^2 } + \omega^2(x)  \sqrt{\rho} =\frac{C^2}{\hbar ^2} \frac{1}{ (\sqrt{\rho})^3}
\qquad\text{\small with}\qquad
\omega ^2(x)= \frac{2m}{\hbar} (E-V(x)) \,.
\een
Similar applications of the Ermakov-Milne-Pinney equation in this context, including those with magnetic fields,
may be found in \cite{Lee,Kaushal}.

\subsection{Bose-Einstein condensates}

Among  the many applications of the Ermakov-Milne-Pinney equation
is one to Bose-Einstein condensates (BEC's) \cite{Herring:2007fw}.

The authors start with the spherically symmetric but time-dependent
Gross-Pitaevski equation in $d$ spatial dimensions with radial coordinate $r$,
\ben
i\,\frac{\p \Psi(r,t)}{\p t}= -\half \nabla ^2 \Psi (r,t) +
\Bigl(\lambda(t)r^2 + \nu(t) |\Psi(r,t)|^2 \Bigr )
\Psi (r,t) \label{GP}  \,,
\een
which may be derived from the Hamiltonian
\ben
H = \half \int \Bigl ( \big | \nabla \Psi   \big |^2
+ \nu(t) \big | \Psi(r,t) \big | ^4 + 2 \lambda(t)
r^2 \big | \Psi(r,t) \big | ^2 |  \Bigr ) r^{d-1} dr \,.
\een
Then, using standard ``moment methods'' in the literature,
an equation is obtained for a quantity $y(t)$ \cite{Herring:2007fw}
\ben
\frac{d}{dt} \Bigl ( y \ddot y   -\half \dot y^2 -2K \nu(t) + 4
\lambda (t) y^2 \Bigr ) = 2 (d-2)K y \frac{d}{dt}\bigl(\frac{\nu(t)}{y}\bigr)
 \een
with $K$ a constant.
The case $d=2$ is evidently special and in that case, setting
$f=\sqrt{y(t)}$ and $\omega^2(t)=2\lambda(t)$,
one obtains (\ref{Dingle2}), with
\ben
\Omega^2 (\tau)= \half C + K \nu(t)\,.
\een

\section{The Eisenhart-Duval Lift and the
  Ermakov-Milne-Pinney Equation}\label{EDliftsec}

\subsection{Eisenhart-Duval lift of Newtonian mechanics}\label{ELCMS}

As originally formulated in \cite{Eisenhart,Bargmann}, the Eisenhart-Duval lift provides a geometric description of a mechanical system with $d$ degrees of freedom $x_i$, $i=1,\dots,d$, and the potential energy $U({t},x)$ in terms of geodesics of the Lorentzian metric on a $(d+2)$--dimensional spacetime
\ben\label{MetriC}
g_{\mu\nu}(y) dy^\mu dy^\nu=-2 U({t},x) d t^2-dt dv+dx_i dx_i,
\een
where $y^\mu=(t,v,x_i)$.
Computing the Christoffel symbols and analyzing the geodesic equations one finds that
$t$ is affinely related to the proper time $\tau$
\ben\label{lt}
\frac{d^2t}{d \tau^2}=0 \quad \Rightarrow \quad \frac{d t}{d \tau}=\kappa,
\een
where $\kappa$ is a constant, $x_i$ obeys Newton's equation (passing from $\tau$ to $t$)

\ben\label{EQN}
\frac{d^2 x_i}{d t^2}+\partial_i U({t},x)=0,
\een
while the dynamics of $v$ is fixed from the condition that the geodesic is null or time--like
\ben\label{eqs1}
\frac{d v}{d t}=\frac{d x_i}{dt} \frac{d x_i}{dt}-2 U-\frac{\epsilon}{\kappa^2},
\een
where $\epsilon=0$ is for null geodesics and $\epsilon=-1$ for time--like geodesics.
Newtonian mechanics is thus recovered by implementing the null reduction along $v$ \cite{Eisenhart}. Note that the original construction \cite{Eisenhart,Bargmann} dealt with null geodesics only.

The spacetime (\ref{MetriC}) belongs to the Kundt class as it admits the covariantly constant null Killing vector field
\ben\label{Xi}
\xi^\mu \partial_\mu=\partial_v, \qquad \nabla_\mu \xi_\nu=0, \qquad \xi^2=0.
\een
The latter can be used to construct the trace--free energy--momentum tensor in a geometrically rather appealing way\footnote{The factor $\frac{d}{2\pi}$, $d$ being the dimension of the $x$--subspace, is chosen for further convenience.}
\ben\label{EMT}
T_{\mu\nu}=\frac{d}{2\pi} \Omega(y)^2 \xi_\mu \xi_\nu, \qquad {T^\mu}_\mu=0, \qquad
\nabla_\rho T_{\mu\nu}=\frac{d}{\pi} \Omega\, \partial_\rho \Omega\,\xi_\mu \xi_\nu,
\een
where $\Omega(y)^2$ is an arbitrary function (the energy density).
The only non--vanishing component of $\xi_\mu$ is $\xi_t=-\frac 12$, which gives $T_{tt}=\frac{d }{8\pi} \Omega^2$ while the rest vanishes.
Since $\xi^\mu \partial_\mu \Omega=0$ holds true if $\Omega$ does not depend on $v$, the energy--momentum tensor is conserved,
\ben
\nabla^\mu T_{\mu\nu}=0,
\een
provided $\Omega=\Omega(t,x)$.

The Ricci tensor which derives from (\ref{MetriC}) has only one nonzero component while the scalar curvature vanishes,
\ben
R_{tt}=\partial_i \partial_i U, \qquad R=0.
\een

Given the Eisenhart-Duval metric (\ref{MetriC}) and the energy--momentum tensor (\ref{EMT}), the Einstein equations
\ben\label{EE}
R_{\mu\nu}-\frac 12 g_{\mu\nu}(R+2\Lambda)=8\pi T_{\mu\nu}
\een
imply that the contribution of the cosmological term necessarily vanishes, $\Lambda=0$, thus reducing (\ref{EE}) to
\ben\label{EE1}
R_{\mu\nu}=8\pi T_{\mu\nu}.
\een
Only the $(tt)$--component is non--trivial; it links $\Omega$ to $U$, as
\ben
\Omega^2=\frac{1}{d} \partial_i \partial_i U.
\een

A particularly interesting example of the Eisenhart-Duval geometry occurs if $U$ and $\Omega$ are $t$-independent. Setting \ben
\Omega^2(x)=\frac{4\pi G}{d} \rho(x)
\een
 and interpreting $G$ as Newton's constant and $\rho(x)$ as the mass density, one recovers the Newton equation for the gravitational potential.

In conclusion, Newtonian mechanics can be represented in terms of a metric which solves the Einstein equations (\ref{EE1}) provided the energy--momentum tensor is chosen in the form (\ref{EMT}).

\subsection{Lifting the Ermakov-Lewis invariant}\label{LewisSect}

The aim of this subsection is to show how applying the Eisenhart-Duval lift to a time-dependent harmonic oscillator allows one to obtain the Ermakov-Lewis invariant by performing a conformal transformation of the metric \cite{Eisenhart,Bargmann}.

The equations of motion of a time-dependent harmonic oscillator follow from the Hamiltonian
\ben
H= \half \Bigl ( m^{-1} (t) p_q^2   + m(t)  \omega^2(t) q^2   \Bigr )
\label{Ham} \,.
\een
Let $H=-p_t $ and let
\ben
{\cal H}(x^\mu,p_\mu) = 2 p_t p_v + m^{-1} p_q^2 +
m \omega^2 q^2 p_v^2 = g^{\mu \nu}p_\mu p_\nu
\een
with $x^\mu = (t,v,q)$, $p_\mu=(p_t,p_v,p_q)$ .  The Eisenhart-Duval metric
on the  extended (``Bargmann" \cite{Bargmann}) spacetime $M=\{\mathbb{R}^3, g \}$ is given by,
\ben
g_{\mu \nu }dx^\mu dx^\nu = 2 dt dv +  m  dq^2   -  m \omega ^2  q^2 d t^2
\label{metric} \,.
\een

The null geodesics, considered as unparametrised
 curves, of a Lorentzian  metric $g_{\mu \nu}$ are the same as for any conformally related Lorentzian metric. They are given by the Hamiltonian
flow on $T^\star M$ with Hamiltonian
\ben
{\cal H} = g^{\mu \nu} p_\mu p_\nu
\label{flow}
\een
subject to the constraint
\ben
{\cal H}=0\,.
\label{con}
\een
 Hamilton's
 equations following from  the Hamiltonian  (\ref{flow})
 with constraint (\ref{con}) for two conformally related metrics
 $g_{\mu \nu}$ and $\Omega^2(x) g_{\mu \nu}$  are identical.

The vector field $V= \frac{\p}{\p v}$ is a null Killing vector field
and thus
\ben
p_v= {\rm constant}.
\een

To obtain the motion in 2 spacetime dimensions we perform a Marsden-Weinstein reduction
(referred to as ``ignoration of  cyclic coordinates'' in old fashioned books)
 bearing in mind the constraint (\ref{con}) and setting $p_v=1$
to recover the Hamiltonian system (\ref{Ham}).
We now set
\ben
Q=\frac{q}{f}
\een
for some function  $f(t)$ to be determined.
In $(Q,t,v)$ coordinates the metric may be cast in the form
\ben
ds^2 = 2 dt(dv + A) + mf^2 dQ^2 - Q^2 g(t)  dt^2,
\een
where $A$ is the one-form
\ben
A = mf\dot f Q dQ + \half mQ^2 (\dot f^2 -f^2 \omega^2) dt  +
\half Q^2   g(t)  dt
\een
and  $g(t)$ is an arbitrary function of time.
If $A$ is closed, $dA=0$, then we may define a new coordinate $\tilde v$ by
\ben
d \tilde v = dv + A,
\een
in terms of which  the metric becomes
\ben
ds ^2 = 2 dt d \tilde v + mf^2 dQ^2 - Q^2 g^2 dt ^2.
\een
Now introducing  a new time coordinate $\tau$ by
\ben
dt = mf^2 d\tau\,,
\een
we have
\ben
ds ^2 = mf^2 \Bigl \{  2 d \tau d \tilde v + dQ^2  - mf^2 Q^2 g d \tau  ^2  \Bigr  \} \,.
\een
The condition for $A$ to be closed  is
\ben
\frac{d}{dt}(m \dot f) + m\omega^2 f = \frac{g}{f}\,.
\een
If
\ben
g = \frac{\Omega^2 }{mf^2}\,,
\een
where $\Omega $ an arbitrary constant,
then the metric becomes
 \ben
ds^2 = mf^2 \Bigl\{  2 d \tau d \tilde v + dQ^2  -  Q^2  d \tau  ^2  \Bigr \} \,.
\label{brace}
\een
As noted above, two conformally related metrics have the same null geodesics, although their affine parameters differ.
From this we deduce that the null  geodesics of the static metric inside the braces
in (\ref{brace}) are the same as those for our  time dependent metric (\ref{metric}).

If we now perform a null reduction on the Killing vector
$\frac{\p}{\p\tilde v}$, we obtain the  Hamiltonian of a time-independent harmonic oscillator
\ben
\tilde H = \half (p_Q^2 + \Omega^2 Q^2).
\een

\subsection{Ermakov-Milne-Pinney cosmology}\label{EMPEDliftsec}

Now we show that the Ermakov-Milne-Pinney equation naturally arises if one incorporates a cosmic scale factor into a suitably chosen Bargmann metric, when the energy--momentum tensor is chosen in a proper way.

Consider a $(d+2)$--dimensional spacetime parametrized by the coordinates $y^\mu=(t,v,x_i)$, $i=1,\dots,d$, and endowed with the Lorentzian metric \cite{Galajinsky}
\ben
\label{EM}
ds^2=-\frac{\gamma^2 x_i x_i }{a(t)^2} dt^2-dt dv+a(t)^2 dx_i dx_i,
\een
where $a(t)$ is an arbitrary function and $\gamma$ is a constant.
For a fixed value of $t$ the line element in the $d$--dimensional slice parametrized by $x_i$ is given by ${a(t)}^2 dx_i dx_i$. Therefore ${a(t)}^2$ may be interpreted as a \emph{cosmic scale factor}. The  metric (\ref{EM}) possesses, as does
its conventional counterpart (\ref{MetriC}),  a covariantly constant null Killing vector field, namely (\ref{Xi}). The choice of (\ref{EM}) will be justified {\rm a posteriori} in sect. \ref{FLRWsec}.

Constructing the energy-momentum tensor following the prescription (\ref{EMT}) and specifying to the class of energy densities which depend on $t$ only, $\Omega=\Omega(t)$, from Eq. (\ref{EE1}) one finds
\ben\label{EP}
\ddot a+\Omega^2(t) a=\frac{\gamma^2}{a^3}\,.
\een
Thus the dynamics of (\ref{EM}) is governed by the Ermakov-Milne-Pinney equation. The instance of\, $\Omega=\mbox{const}$ has been discussed recently in \cite{Galajinsky}, in which case the Ermakov-Milne-Pinney equation reduces to conformal mechanics in one dimension \cite{DFF}.

One can learn more about the geometry of (\ref{EM}) by analyzing the geodesic equations. Computing the Christoffel symbols, one concludes that $t$ is affinely related to the proper time,
\ben
\frac{d^2 t}{d \tau^2}=0\, .
\een
The coordinate $x_i$ obeys in turn the oscillator--like equation
\bea\label{Geo}
a^2 \frac{d}{dt} \left(a^2 \frac{d}{dt} x_i \right)+\gamma^2 x_i=0,
\eea
in which we passed from $\tau$ to $t$, while the evolution of $v$ is fixed from the condition that the geodesic be null or time--like. Eq. (\ref{Geo}) prompts one to introduce the conformal time
\ben\label{CT}
a^2(t) \frac{d}{dt}=\frac{d}{d\eta}\,, \qquad \eta(t)=\int_{t_0}^t \frac{d\tilde t}{a^2(\tilde t)}\,,
\een
which brings the metric (\ref{EM}) to the form
\ben\label{EM1}
ds^2=a^2(\eta)\left(-\gamma^2 x_i x_i d\eta^2-d\eta dv+dx_i dx_i\right),
\een
where $a(\eta)=a(t(\eta))$, and $t(\eta)$ is the inverse of $\eta(t)$ in (\ref{CT}). Eq. (\ref{EM1}) is an analog of the flat ($k$=0) FLRW cosmological model in which the Minkowski metric has been changed into the simplest PP--wave; the Friedmann equation which determines the evolution of the cosmic scale factor is replaced, in this case, by the Ermakov-Milne-Pinney equation.

To conclude this section we note that the coordinate transformation
\ben\label{TRF}
t'=t, \qquad x'_i=a(t) x_i, \qquad v'=v+a(t) \dot a(t) x_i x_i
\een
brings (\ref{EM}) to the form
\ben
ds^2= \left(\frac{\ddot a}{a}-\frac{\gamma^2}{a^4} \right) x_i x_i dt^2-dt dv+dx_i dx_i,
\een
where we omitted the primes,
or, in view of (\ref{EP}),
\ben
\label{Ometric}
\mybox{
ds^2= -\Omega(t)^2 x_i x_i dt^2-dt dv+dx_i dx_i}
\een
which is the $d+2$ dimensional Bargmann metric associated with an isotropic oscillator in $(d,1)$ dimensions with time--dependent frequency $\Omega(t)$,
the  motions of which  are the projections of the null geodesics of (\ref{Ometric}).

\subsection{Symmetries as conformal Killing isometries}\label{ConfKillingSec}

Finding the symmetries of time-dependent harmonic oscillators generated an extensive literature in the early 1980's, including \cite{Ray:1980se,Ray:1980kw,RayReid2,PrinceEliezer}.
A discussion in terms of canonical transformations
is in \cite{Pedrosa}. An alternative approach is presented below in terms of the Eisenhart-Duval lift. For  simplicity, we stick to Eq. (\ref{Ometric}).

Following  \cite{Bargmann},
the symmetries of a non-relativistic system in $(d,1)$ dimensions can be obtained as a subgroup of the conformal symmetries of the $d+2$ dimensional Bargmann manifold: one selects those conformal transformations that leave invariant the covariantly constant null vector $\p_v$.

In our case, the symmetries given in Ref. \cite{PrinceEliezer} are seen to be consistent with the Schr\"odinger group in $d$ dimensions -- which is a subgroup of the conformal group of the extended (Bargmann) spacetime. Now we re-derive the above-mentioned symmetries  in the specific case of our Ermakov-Milne-Pinney spacetime of sec.\ref{EMPEDliftsec}.

Consider indeed a generic infinitesimal transformation
\ben\label{transf}
t'=t+\delta t, \qquad  v'=v+\delta v, \qquad x'_i=x_i+\delta x_i\,,
\een
where $\delta t$, $\delta v$, $\delta x_i$ are arbitrary functions of $t,v,x$.
Demanding (\ref{Ometric}) to be invariant under (\ref{transf}) up to a conformal factor,
\ben
ds'^2=(1+\Lambda) ds^2,
\een
one gets a coupled set of partial differential equations to fix $\delta t$, $\delta v$, $\delta x_i$ and $\Lambda$. Omitting details, we present the (conformal) isometries of (\ref{Ometric}),
\besub
\begin{align}
&
t'=t+\lambda(t)+\frac 12 \epsilon x^2-x_i \rho_i(t),
\\[2pt]
&
x'_i=x_i+\mu_i (t)+\omega_{ij} x_j+\frac 12 \kappa x_i +\frac 12 \dot\lambda (t) x_i+\frac 12 v \epsilon x_i
\\[2pt]
&
\qquad
+\frac 12 x^2 \dot{\rho}_i (t)-x_i x_j \dot{\rho}_j (t)-\frac 12 v \rho_i(t),
\\[2pt]
&
v'=v+\nu+2{\dot\mu}_i(t) x_i+\frac 12  \ddot\lambda (t) x^2-\frac 12 \epsilon\, \Omega^2x^2 x^2 +\kappa v+\frac 12 \epsilon v^2
\\[2pt]
&
\qquad  +\Omega^2 x^2 x_i \rho_i(t)-v x_i \dot{\rho}_i (t),
\end{align}
\label{isomeries}
\esub
where $\epsilon$, $\nu$, $\kappa$, $\omega_{ij}=-\omega_{ji}$ are constant infinitesimal parameters, $x^2=x_i x_i$, the functions $\mu_i(t)$, $\rho_i(t)$, and $\lambda(t)$ obey the ordinary differential equations
\ben\label{constr}
\ddot \mu_i+\Omega^2 \mu_i=0, \qquad
\ddot \rho_i+\Omega^2 \rho_i=0, \qquad \lambda^{(3)}+4 \Omega^2\, \dot\lambda+4 \Omega\dot\Omega\, \lambda=0,
\een
while the conformal factor is
\ben\label{CF}
\Lambda=\kappa+\dot\lambda (t)+v \epsilon-2 x_i \dot{\rho}_i (t).
\een
Taking into account the order of the differential equations (\ref{constr}), one finds that the transformations (\ref{isomeries}) involve $6+4d+\frac{d(d-1)}{2}=\frac{(4+d)(3+d)}{2}$ independent infinitesimal parameters
--- the same as that of the flat-space conformal group in $d+2$ dimensions.

From Eq. (\ref{CF}) one concludes that the $\kappa$, $\dot\lambda (t)$, $\epsilon$, and $\rho_i(t)$--transformations give rise to conformal Killing vectors\footnote{Note that a combination of the $\kappa$ and $\lambda$ -transformations may result in a Killing vector field, provided $\dot\lambda=-\kappa$. In view of (\ref{constr}), this only happens if $\Omega(t)={g}/{t}$, where $g$ is a constant.}, while the $\nu$, $\mu_i(t)$, $\omega_{ij}$ and constant $\lambda$--transformations generate Killing vectors. In view of (\ref{constr}), the isometry with constant parameter $\lambda$ is only possible for constant frequency $\Omega$ which corresponds to a stationary spacetime. Time translations are broken in general.

We notice that, while distinct conformal isometries act on the coordinates as in \eqref{isomeries}, the functions $\lambda(t)$, $\rho_i(t)$ and $\mu_i(t)$ are not independent. Obviously, $\rho_i(t)$ differs from $\mu_i(t)$ by an inessential constant. If $\mu$ satisfies the first equation in \eqref{constr}, then $\lambda(t) = \mu^2(t)$ will satisfy the second. Moreover, if $\mu_1$ and $\mu_2$ are independent solutions of the first equation, then the three independent solutions of the second are given by $\mu_1^2$, $\mu_2^2$, $\mu_1 \mu_2$. The deeper reason why this happens is that the conformal Killing vectors form a Lie algebra, which we identify below with $so(2,2+d)$.

The interpretation of the (conformal) Killing vectors above and their algebra becomes more transparent if one switches to conformal time, (\ref{EM1}). It is evident that all (conformal) isometries of the PP-wave metric
\ben
ds^2=-\gamma^2 x_i x_i d\eta^2-d\eta dv+dx_i dx_i,
\label{confPP}
\een
will be automatically transmitted to become (conformal) isometries of (\ref{EM1}), since the two expressions only differ by the cosmic scale factor $a(\eta)$. In particular, all the symmetry transformations of (\ref{confPP}), which do not involve $\eta$ explicitly, will be transformed into the Killing vectors of (\ref{EM1}), while those affecting $\eta$ will be transmitted into conformal Killing vectors.

Our clue is that the metric above is \emph{conformally flat}. This follows from the vanishing of the Weyl tensor. Skipping details we merely mention that it can also be seen, explicitly, by applying the Arnold transformation, see \cite{Guerrero:2013bva,Lopez-Ruiz:2014vpa,Cariglia:2016oft}.

The group of conformal transformations of \emph{any} $(1,1+d)$--dimensional conformally flat spacetime is isomorphic to that of Minkowski space, explaining the ``coincidence'' we noted earlier.

The explicit form of the (conformal) isometries of (\ref{confPP}) follows from Eqs. (\ref{isomeries}), (\ref{constr}) at $\Omega=1$ after the  substitution $t \to \eta$. In particular, conformal time translations, $\eta\to \eta+\theta$, are now symmetries, because $\gamma=\const$ in (\ref{confPP}).

From the first equation in (\ref{constr}) one finds,
\ben
\mu_i(\eta)=\alpha_i \cos{(\eta)}+\beta_i \sin{(\eta)},
\een
where the infinitesimal parameters $\alpha_i$ and $\beta_i$ are associated with spatial translations and Newton--Hooke boosts \cite{GP}. The second equations generate 
\ben
\rho_i(\eta)=\tilde\alpha_i \cos{(\eta)}+\tilde\beta_i \sin{(\eta)},
\een
which involve the infinitesimal parameters $\tilde\alpha_i$, $\tilde\beta_i$ and provide contributions to Eqs. (\ref{isomeries}) which are nonlinear in $x_i$.
The third equation yields
\ben
\lambda(\eta)=\theta+\sigma\cos{(2\eta)}+\rho\sin{(2\eta)},
\een
where the infinitesimal parameters $\theta$, $\sigma$, $\rho$, linked to time translations, special conformal transformations, and dilatations form an $so(2,1)$ subalgebra. Along with spatial rotations described by $\omega_{ij}$, the $\mu_i(\eta)$ and $\lambda(\eta)$--transformations form the
 \emph{conformal Newton--Hooke algebra}. For a detailed discussion of the Schr\"odinger and conformal Newton--Hooke algebras and their realizations in spacetime see e.g. \cite{DHNC,Galajinsky1}.

 As it follows from (\ref{isomeries}), the isometries of (\ref{confPP}) also involve the translation in the $v$--direction, while the set of conformal isometries contains the $\kappa$, and $\epsilon$--transformations.

Note that the $\kappa$--transformation in (\ref{isomeries}) and (\ref{CF})
is realized in a way analogous to  conventional dilatation  in the Schr\"odinger algebra. It derives from the latter by replacing  the temporal variable $t$ by the ``null" coordinate $v$; it has appeared before in the context of gravitational waves \cite{Torre}. The $\epsilon$--transformation is in turn an analog of  special conformal transformation, again $t$ replaced by $v$.  It is straightforward to verify that, along with the translations in the $v$--direction, $v'=v+\nu$, they form an $so(2,1)$ subalgebra. Interestingly enough and extending the Galilei-Carroll ``duality" \cite{GalCarroll}, the latter acts upon the null coordinate $v$ in very much the same way as $so(2,1)$ entering the conformal Newton-Hooke algebra affects the temporal coordinate $t$.

To summarize, the algebra of vector fields which involve both Killing and conformal Killing vectors can be identified with $so(2,2+d)$, the conformal Newton-Hooke algebra being its subalgebra.

Having identified the conformal isometries of the metric (\ref{Ometric}), the symmetries of the underlying classical system in one less dimension (i.e., the time-dependent oscillator) could now be derived. Skipping details, we just mention that implementing the null reduction along $v$, the $SO(2,1)$ conformal subgroup with parameters $\nu$, $\kappa$ and $\epsilon$ in (\ref{isomeries}) will be broken, allowing us to recover the Schr\"odinger symmetry found  in  \cite{PrinceEliezer}.
 The generators are conveniently identified using the formulae in sec. 3 of \cite{HH2000}.

\subsection{Geodesic motion in Ermakov-Milne-Pinney cosmoi}\label{GeoMot}

Having established a link between the Ermakov-Milne-Pinney equation and the Eisenhart-Duval lift, let us study the geodesic motion in Ermakov-Milne-Pinney cosmoi. The analysis is facilitated by switching to the conformal time (\ref{EM1})
which allows one to solve the geodesic equations by quadrature,
\besub
\begin{align}
&
\tau=\kappa \int_{\eta_0}^\eta a^2(\tilde\eta) d\tilde\eta +\tau_0,
\\[2pt]
&
x_i(\eta)=\alpha_i \cos{(\gamma\eta)}+\beta_i \sin{(\gamma\eta)},
\\[2pt]
&
v(\eta)=v_0-\frac 12(\alpha^2-\beta^2) \gamma \sin{(2\gamma\eta)}+\alpha\beta\gamma \cos{(2\gamma\eta)}-\epsilon \kappa^2 \int_{\eta_0}^\eta a^2(\tilde\eta) d\tilde\eta,
\end{align}
\label{GEO}
\esub
\begin{figure}[ht]
\begin{center}
\resizebox{0.38\textwidth}{!}{%
\includegraphics{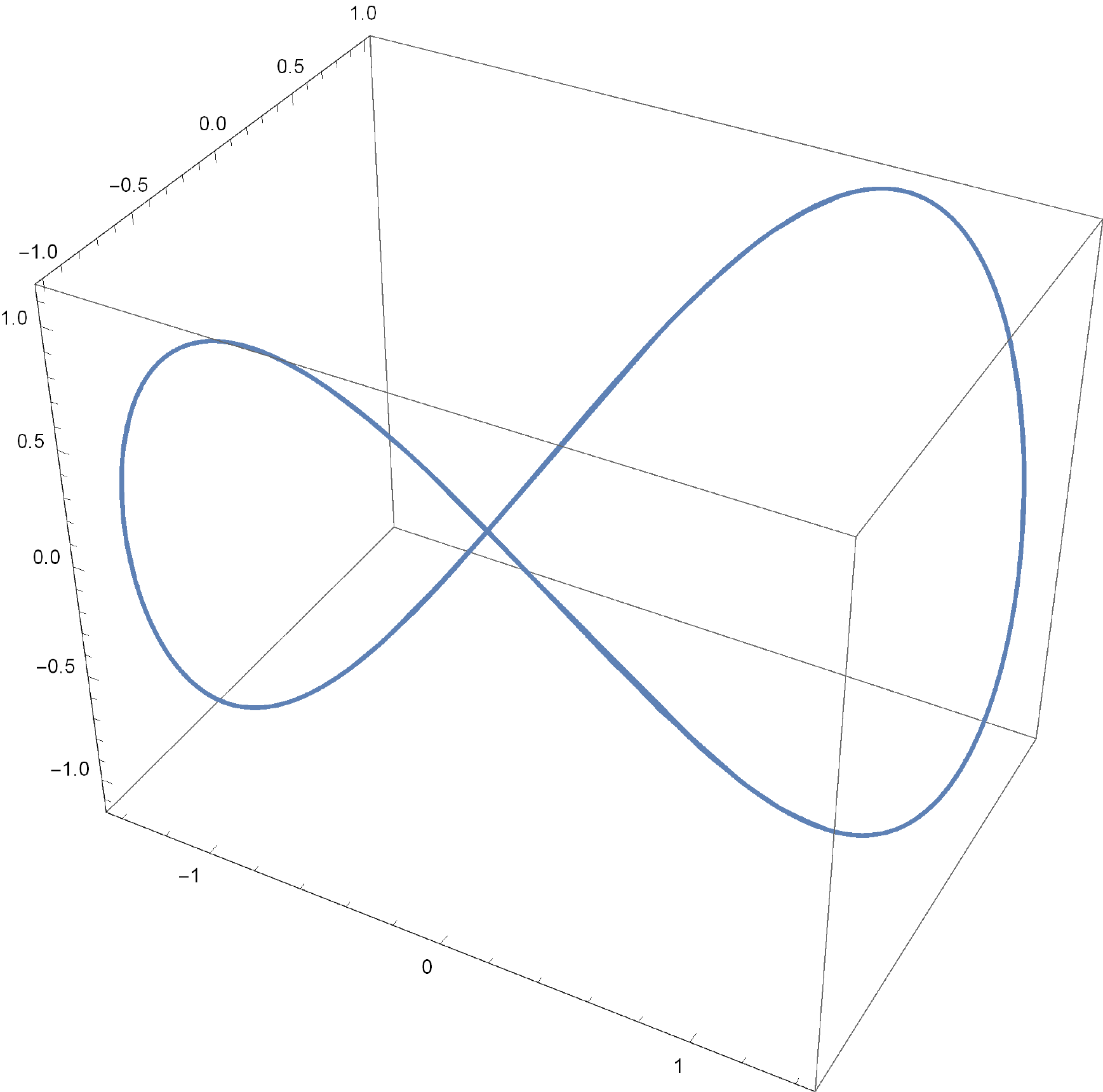}}\vskip-4mm
\caption{\small Null geodesics in Ermakov-Milne-Pinney cosmoi are closed loops which wrap around elliptic cylinders.
The graph lies in three--dimensional space with $(x_1,x_2,v)$ parametrizing the two horizontal and the vertical axes, respectively.}
\label{fig1}
\end{center}
\end{figure}
where $\tau_0$, $\kappa$, $\alpha_i$, $\beta_i$, $v_0$ are constants of integration and $\alpha^2=\alpha_i \alpha_i$, $\alpha\beta=\alpha_i \beta_i$. It is assumed that $\epsilon=0$ for null geodesics and $\epsilon=-1$ for time--like geodesics.

Before proceeding, it is worth mentioning that, since for null geodesics Eqs. (\ref{GEO}) admit a particular solution
\ben
x_i(\eta)=0, \qquad v(\eta)=v_0,
\een
the metric (\ref{EM1}) is formulated in a reference frame comoving with a light signal which travels along the $v$--axis. This correlates with the fact that (\ref{EM1}) differs from a PP--wave by a scale factor only.

The first line in (\ref{GEO}) defines $\eta$ as an implicit function of the proper time $\tau$. Although in most cases of interest the integral $\int_{\eta_0}^\eta a^2(\tilde\eta) d\tilde\eta$ cannot be evaluated exactly, Eqs. (\ref{GEO}) prove to be sufficient to comprehend a qualitative behaviour of geodesics. Indeed, the second line in (\ref{GEO}) defines an ellipse. By making use of rotational invariance, one can set the ellipse to lie, say, in the $x_1 x_2$--plane. As $v$ evolves with time, geodesics in the Ermakov-Milne-Pinney cosmology wrap around the elliptic cylinder, $v$ being its axis.

For null geodesics ($\epsilon=0$), the trajectory is a closed loop and the motion is periodic (see FIG. 1).
For time--like geodesics ($\epsilon=-1$), the orbit wraps around the cylinder remaining in a compact region of space for some time, then the $\epsilon$--term in the last line in (\ref{GEO}) starts dominating and the particle escapes (see FIG. 2). Given the coordinate system in which the metric (\ref{EM1}) is formulated, this happens because massive particles travel slower than a reference frame comoving with a light signal which propagates along the $v$--axis.
\begin{figure}[ht]
\begin{center}
\resizebox{0.38\textwidth}{!}{%
\includegraphics{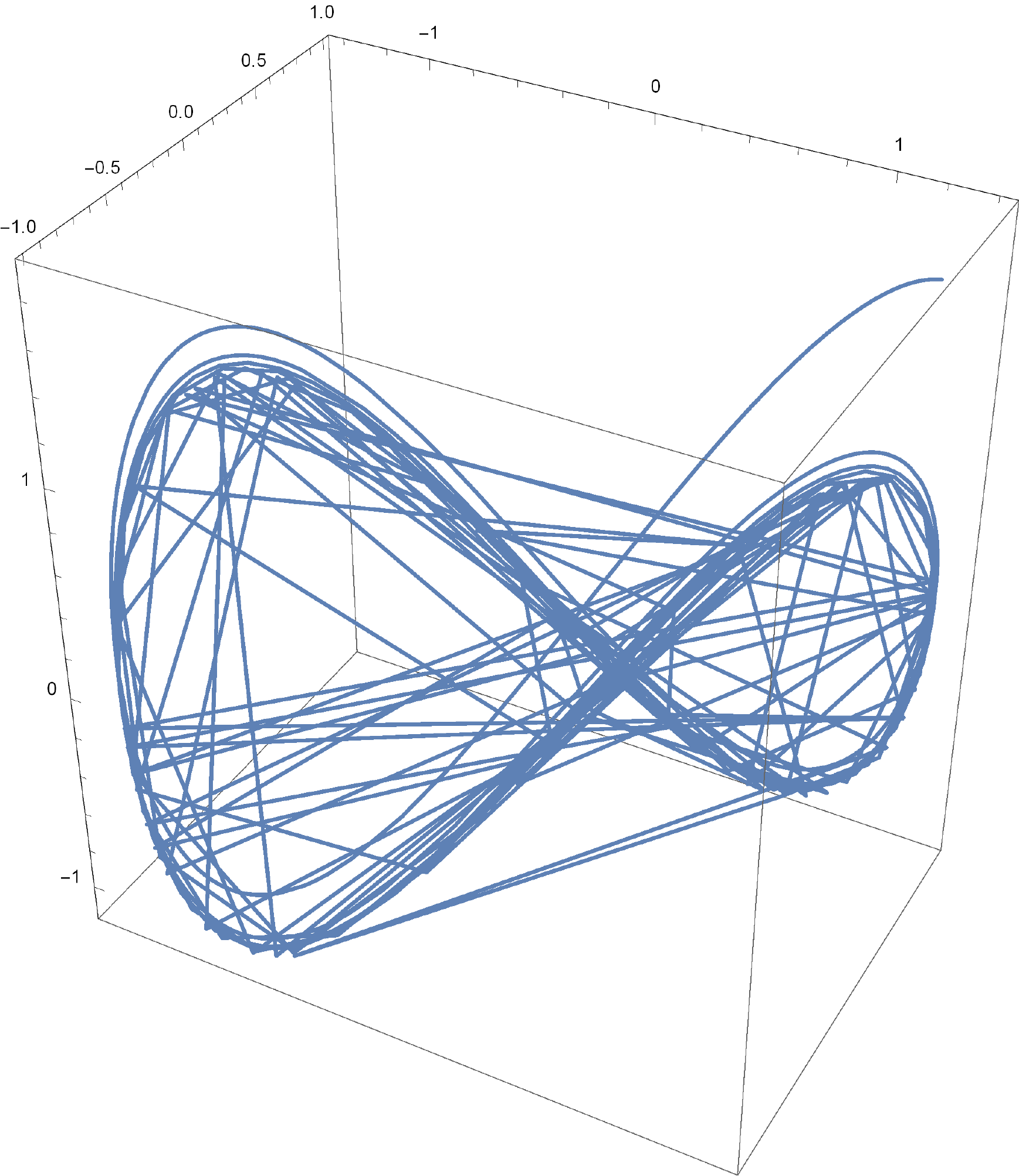}}\vskip-4mm
\caption{\small Time--like geodesics in  Ermakov-Milne-Pinney cosmoi wrap around elliptic cylinders. They remain in a compact region of space for some time before escaping.
The graph is given in three--dimensional space, $(x_1,x_2,v)$ parametrizing the two horizontal and one vertical axes, respectively.}
\label{fig2}
\end{center}
\end{figure}

Little is known about analytic solutions to the Ermakov-Milne-Pinney equation. Assuming that the energy density $\Omega^2(t)$ decreases with time (which happens in an expanding universe),  one has (see, e.g., \cite{Lewis2} and Sect. 2.4.2 in \cite{ZP})\footnote{From now on we switch back to cosmic time $t$.}
\besub
\begin{align}
&
\Omega^2(t)=\frac{h^{n-2}}{4 \nu^2 t^n}, \qquad \nu=\frac{1}{2-n}, \qquad n\ne 2,
\\[2pt]
&
a(t)=\sqrt{\pi\gamma \nu t(J_\nu^2 (\lambda)+Y_\nu^2 (\lambda))}, \qquad \lambda={\left(\frac{t}{h}\right)}^{\frac{1}{2\nu}},
\end{align}
\label{Cosmoi}
\esub
where $n$ is a rational number such that $\nu$ is positive, $h$ is a positive constant, and $J_\nu$, $Y_\nu$ are Bessel functions. Note that in these cases $a(t)$ is a monotonically increasing convex function which tends to a fixed nonzero value as $t\to 0$.

Other interesting examples are provided by negative integer $\nu$; then $\nu$ should be replaced by $|\nu|$ in the expression for $a(t)$ in Eq. (\ref{Cosmoi}). These models are represented by monotonically increasing convex functions starting at $a(0)=0$. The instance $t=0$ can be interpreted as the Big Bang. A typical example is shown in FIG. 3.

A notable simplification takes place for $\ddot a=0$,  when the solution is expressed in terms of elementary functions
\ben
\Omega(t)=\frac{h}{t^2}, \qquad a(t)=\sqrt{\frac{\gamma}{h}}\, t \qquad \Rightarrow \qquad \Omega(t)=\frac{\gamma}{{a(t)}^2}\,.
\een
This case corresponds to a linearly expanding universe in which the energy density of matter decreases consistently with the inverse square law. The general solution to the geodesic equations is,
\begin{figure}[ht]
\begin{center}
\resizebox{0.38\textwidth}{!}{%
\includegraphics{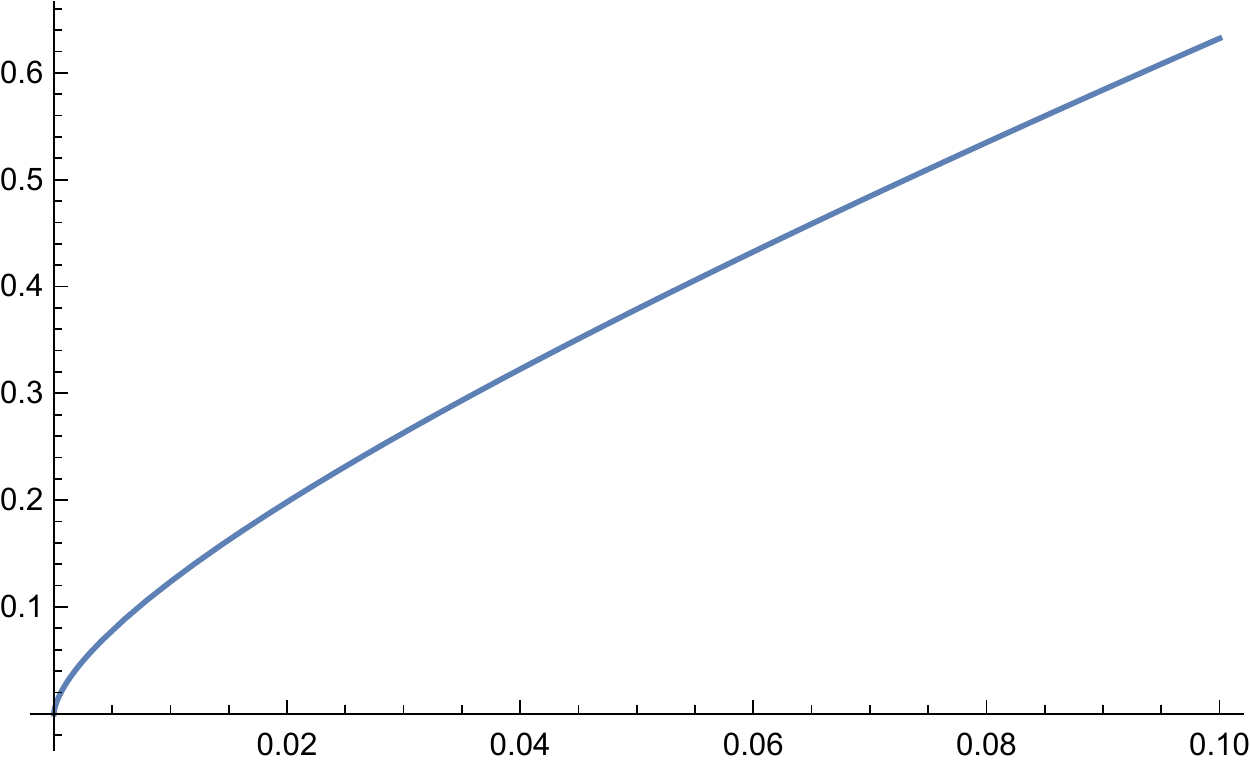}}\vskip-4mm
\caption{\small A graph of the cosmic scale factor in the Ermakov-Milne-Pinney cosmological model at $n=3$.}
\label{fig3}
\end{center}
\end{figure}
\besub
\begin{align}
&
t=t_0+\tau/\kappa,
\\[2pt]
&
x_i(\tau)=\alpha_i \cos{\left( h/t\right)}+\beta_i \sin{\left( h/t\right)},
\\[2pt]
&
v(\tau)=v_0-\epsilon \kappa^2 t- \alpha \beta \gamma \cos{\left(2h/t\right)}+\frac{1}{2}(\alpha^2-\beta^2) \gamma \sin{\left(2h/t\right)},
\end{align}
\label{or1}
\esub
where $t_0$, $\kappa$, $\alpha_i$, $\beta_i$, $v_0$ are constants of integration.

Although the graphs representing the cosmic scale factors in the Ermakov-Milne-Pinney cosmology look quite reasonable, the geodesic motion is apparently unrealistic. This happens because the second line in (\ref{GEO}) defines an ellipse. An obvious cure is to generalize the construction to the case of time dependent $\gamma$ which will alter the qualitative behaviour of geodesics.
It proves sufficient to replace $\gamma$ in (\ref{EM}) by an arbitrary function $\gamma(t)$ which will then show up on the right hand side of Eq. (\ref{EP}),
 viewed as an \textit{algebraic equation} to fix $\gamma(t)$ in terms of $a(t)$ and $\Omega(t)$. In this way one can model a reasonable geodesic behavior in the generalized Ermakov-Milne-Pinney cosmoi by properly choosing the cosmic scale factor and the energy density. Note, however, that, as the associated geodesic equations involve a time dependent oscillator, with frequency $\gamma(t)^2$, finding an analytic solution may  be complicated.

\vspace{0.5cm}

\section{Friedmann-Lema\^\i tre-Robertson-Walker Spacetimes}\label{FLRWsec}

There has been considerable interest in the past few years
in applying the ideas circling around the Ermakov-Milne-Pinney equation
to problems in cosmology, both classical and quantum,
\cite{Rosu:1998wr,Rosu:1999hp,Rosu:1999ud,
Rosu:1999rs,Rosu-Espinoza,Hawkins:2001zx,Williams:2003vb,Graham:2011nb}.
In this section we shall establish some connections. Before doing so we shall begin by establishing our notations and conventions.

A Friedmann-Lema\^\i tre-Robertson-Walker (FLRW) spacetime is one carrying the metric
\ben
g_{\mu \nu}dx^\mu dx^\nu = - dt ^2 + a^2(t)\, d {\Sigma^2}_K
\,,\een
where $d{\Sigma^2}_K$ is a 3-dimensional metric   of
constant curvature $K$.  The {\it scale factor}  $a(t)$ is governed by
three equations of which, if $\dot a \ne 0$, only two
are independent. These are the {\it Raychaudhuri equation}
\ben
\frac{\ddot a}{a} = -\frac{4 \pi G}{3}  (\rho +3P)  + \frac{\Lambda}{3}  \,, \label{Ray}
\een
the  {\it Friedmann  equation}
\ben
{\left(\frac{\dot a}{a} \right)}^2 + \frac{K}{a^2}  = \frac{8\pi G}{3}\rho  + \frac{\Lambda}{3} \,, 
\label{Fried}
\een
and the {\it conservation equation}
\ben
\dot \rho + 3\frac{ \dot a}{a} ( \rho + P) =0 \,,
\label{coneqn}\een
where $\rho$ is the energy density,
$P$ the pressure and $\Lambda$ the cosmological constant.
\goodbreak

\subsection{Matter models}\label{MattMod}

The simplest models for the matter content  are those in which it consists
of a perfect fluid with constant barotropic index.
This means that
\ben
P=(\gamma-1) \rho = w \rho\,,
\een
where $\gamma$, or equivalently $w$, is constant.
It follows from (\ref{coneqn})  that
\ben
\rho = \frac{\rho_0}{a^{3 \gamma}} \,.
\een
One has the following special cases
\begin{itemize}
\item  $\gamma=1$ corresponds to pressure--free matter.
\item $\gamma =\frac{4}{3}$ to radiation
\item $\gamma =0$ to a  cosmological constant
  which is equivalent to a constant energy density
$\rho= \frac{\Lambda}{8\pi G}$.
\item  $\gamma=2 $ is the largest value of $\gamma$ consistent
  with the dominant energy condition. If the energy density is positive it is  sometimes known as ``stiff matter''. If $\gamma >2$ there exist solutions for which the scale  factor
  blows up in finite time.
\item $\gamma = \frac{2}{3}$ corresponds to a gas of cosmic strings
\item $\gamma = \frac{1}{3}$  corresponds to a gas of membranes
\item  If $\gamma < \frac{2}{3}$, there exist solutions exhibiting a ``big rip'', that is for which the scale factor blows up in finite time \cite{Caldwell:2003vq}.

\end{itemize}

Another simple model is to suppose that the matter consists of a scalar
field $\phi$ with potential $V(\phi)$
\ben
\rho= \half \dot \phi^2 + V(\phi) \,, \quad  P= \half \dot \phi^2 -V(\phi)\,.
\een
It follows from (\ref{coneqn}) that
\ben
\ddot \phi + 3 \frac{\dot a}{a} \dot \phi + V^\prime(\phi) =0\,.
\een
If $|V(\phi)| <<  \half \dot \phi^2 $ this behaves like stiff matter, $\gamma=2$. If $|V(\phi)| >>  \half \dot \phi^2$
this behaves like a cosmological constant.

\subsection{Cosmic clocks and temporal diffeomorphisms} \label{CoClocks}

The coordinate $t$ is called {\it cosmic time} and may be  measured by atomic clocks.
To good approximation it also coincides with local astronomical time.
For some purposes it is convenient to introduce another coordinate $\eta$ called {\it conformal time}, defined  by
\ben
\eta = \int \frac{dt}{a(t)} \,,
\een
in terms of which the FLRW metric is manifestly conformally static
\ben
ds ^2 = a^2(t) \Bigl\{-d \eta ^2 + d {\Sigma^2_K}          \Bigr \}\, .
\een
If $\chi$ is radial distance, conformal time may be measured by light clocks. In other words,
the ticks of a light clock correspond to light rays being bounced between two co-moving mirrors. This procedure is often attributed to Marzke
and Wheeler \cite{Marzke}, but in fact it goes back to Einstein, Robb, Milne, Whitrow and Walker. A survey of explicit solutions of the Friedmann equations
for  matter with a constant barotropic index both as a function of cosmic time and of conformal time may be found in \cite{Harada:2018ikn}.

Concerns about the notion of a  Big Bang and the   beginning of the universe
have led to consideration of other  cosmic clocks or choices of
what have variously been called \emph{time scales} or \emph{time graduations}
\cite{Misner:1970yt,Leblond1,Leblond2}. One obvious choice is the temperature or equivalently the redshift
\ben
T=\frac{a(t_0)}{a(t)} T_0 = T_0(1+z) \,,
\een
where $T_0\approx 3K$ and the subscript $0$ denotes the present epoch.
Now if  $a(t)$ runs from $0$ at the Big Bang to $\infty$
in the infinite future then $T$ runs from $\infty$ to $0$. In fact
given three successive instants of time $t_1 \prec t_2 \prec t_3 $ the
temperature  changes,
$T_{12} = \frac{T (t_1)} {T(t_2)} $ etc  behave multiplicatively
\ben
T_{13} = T_{12} T_{23} \,,
\een
but taking logarithms $T= e^{-\Omega} $  gives an additive scale
\cite{Leblond1,Leblond2}:
\ben
\Omega_{13}= \Omega _{12}+ \Omega_{23} \,,
\een
which runs from $-\infty$ to $+ \infty$. Since from (\ref{Fried}) we have
\ben
d\Omega = \frac{da}{a} = \frac{\dot a}{a} dt
\een
the FLRW metric takes the Eternal  Peripatetic  Form
\ben
g_{\mu \nu}dx^\mu dx^\nu = -
\frac{ d \Omega ^2} { \frac{8 \pi G \rho  }{3}   +
\frac{\Lambda}{3} - K^2 e^{-2 \Omega}}  +
e^{2 \Omega} d {\Sigma^2_K}
\,.
\een

Of course merely introducing  a new time coordinate will not turn a geodesically incomplete spacetime into a geodesically complete one,
but it does underscore the point that rejecting a geodesically
incomplete spacetime as being singular requires a physical
justification, not merely a convenient mathematical definition in order to  prove  theorems.

For a perfect fluid with constant barotropic index
and vanishing cosmological constant,
another temporal diffeomorphism defined by
\ben
d\tau = \frac{2}{3 \gamma-2}{2a} dt
\een
combined with the definition
\ben
x=\sqrt{\frac{3}{8 \pi G \rho_0} } a^{\frac{3 \gamma -2}{2}} \, \quad \Longleftrightarrow
\quad a= \Bigl(\frac{8 \pi G \rho_0}{3} x^2 \Bigr) ^{\frac{1}{3 \gamma-2}}
\een
reduces the Friedmann  equation  (\ref{Fried}) to that
of a time-independent harmonic oscillator
\ben
\bigl(\frac{dx}{d\tau}\bigr)^2 + Kx^2 = 1 \,.
\een

If $K>0$ this is a simple harmonic oscillator. If
$K<0$ this is an  ``inverted'' simple harmonic oscillator.
Both  may be construed as a free particle moving in
a 2-dimensional Newton-Hooke spacetime with coordinates $(\tau,x)$ \cite{GP}.
The systems are invariant under the conformal extensions of Newton--Hooke-groups $NH_\pm$  \cite{GP,Galajinsky1}.

A related construction works if $K=0$ but we include a cosmological
constant term or dark energy term.

If $\Lambda>0$ multiplication of  (\ref{Fried}) by $a^{3\gamma}$ leads
to an upside down  simple harmonic oscillator equation for $a^{\frac{3 \gamma}{2}} $
with solutions of the form
\ben
a= \Bigl [ \sqrt{ \frac{8 \pi G \rho_0}{\Lambda }} \sinh \bigl( \frac{3 \gamma}{2} \sqrt{\frac{\Lambda}{3}} t   \bigr  )       \Bigr ] ^{\frac{2}{3 \gamma}}.
\label{solution}\een

One may calculate the jerk \cite{Dunajski:2008tg}
\ben
j= \frac{a^2 \frac{d^3 a}{dt ^3}}{ \bigl( \frac{da}{dt} \bigr )^3}
= 1 + \frac{9 \gamma (\gamma-1)}
{ 2 \cosh^2 ( \frac{3 \gamma}{2} \sqrt{\frac{\Lambda}{3}} t  ) } \,.
\een
If $\gamma=1$, $j=1$  for all times. Otherwise it merely starts from
a value greater than one and monotonically decreases to
1 at late times. For example  for radiation, $\gamma=\frac{4}{3}$, the jerk
starts from $3$.

If $\Lambda <0$ the motion is converted to a standard harmonic
oscillator. For radiation  $\gamma =\frac{4}{3}$
and  (\ref{Ray}) becomes
\ben
\ddot a + \frac{|\Lambda|}{3} a= -
  \frac{8 \pi G \rho_0}{a^3 }\,, \label{Pinney}
\een
which is a variant of the Ermakov-Milne-Pinney equation \cite{Hawkins:2001zx}.
Equation (\ref{solution}) gives the particular solution:
\bea
a=
\Bigl[\sqrt{ \frac{8 \pi G \rho_0}{|\Lambda|}}
  \sin \bigl(2 \sqrt{\frac{|\Lambda|}{3}} t \bigr) \Bigr]^{\half}
=
\Bigl [\sqrt{ \frac{8 \pi G \rho_0}{-\Lambda }}
 2  \sin \bigl( \sqrt{\frac{|\Lambda|}{3}} t  \bigr)   \cos \bigl( \sqrt{\frac{|\Lambda|}{3}} t \bigr)       \Bigr ] ^{\half}\,.
\label{radiation}
\eea
One may check that the solution  (\ref{radiation}) is a special case
of the general solution given by (\ref{PRO1}) and (\ref{PRO2}).

The instance  $\gamma = \frac{2}{3}$ or $w=\gamma-1 = -\frac{1}{3} $ and $
\Lambda <0$ gives the special $k=0$ case of the
so-called simple harmonic universe \cite{Graham:2011nb}. It takes the form
\ben
a = \frac{4 \pi G \rho_0}{|\Lambda|} \Bigl( 1- \cos \bigl(  \sqrt
{\frac{| \Lambda|}{3}} t \bigr )  \Bigr)\,.
\een

The general solution of (\ref{Ray}) subject to the constraint
(\ref{Fried}) and an origin of time is
\ben
a = \frac{4 \pi G \rho_0}{|\Lambda|} -
\sqrt{
\bigl (\frac{4 \pi G \rho_0}{|\Lambda |}  \bigr ) ^2 - \frac{3K}{|\Lambda |}
}
\cos \bigl(\sqrt{\frac{|\Lambda|}{3}}  t \bigr)\,.
\een
If $K=0$ this coincides with our previous solution;
if $K>0$ it is the non-singular cyclic universe of \cite{Graham:2011nb} .

For more examples of solutions of the Friedmann and Raychaudhuri equations
the reader is referred to \cite{Chen:2014fqa,Chen:2015kza,Chen:2015cja}.

\subsection{Lifting the Friedmann-Lema\^\i tre-Robertson-Walker geodesics}\label{EisenFLRWSec}

Having presented the main features of FLRW cosmologies, now we show how the Eisenhart-Duval lift can be applied to geodesic motion in FLRW spacetimes and how to derive its dynamical symmetries. We then extend the result to the Dmitriev-Zel'dovich equations.

Our starting point is the geodesic Hamiltonian
\ben
H_{FLRW} = \frac{1}{2a^2} \left( - p_\eta^2 + h^{ij} p_i p_j \right)\,,
\label{FLRWHam}
\een
where $\eta$ is the conformal time introduced in the previous section, and $h_{ij}$, $i, j, = 1, \dots 3$ is the metric on the 3--dimensional space.
Setting $H_{FLRW} = - {m^2}/{2}$ and solving for $p_\eta$, we get~:
\ben
p_\eta = \sqrt{m^2 a^2(\eta)+ h^{ij} p_i p_j} \,\, .
\een
We identify the low-energy regime as $\frac{p_i p_i}{m^2 a^2(\eta)} << 1$. In an expanding universe $p_i$ is constant. Therefore the dynamics
will, unlike in the usual non-relativistic case, always enter this regime for  large enough $\eta$.
 In this limit
\ben
p_\eta = m a(\eta)  + \frac{h^{ij} p_i p_j}{2ma(\eta)} \, ,
\label{eq:relationship_nonhom}
\een
which is a relationship between dynamical variables  of order 2 in the momenta.
Then an important step is to transform this relation into a homogeneous one by introducing a conserved momentum, $p_v$  --- which is the physical equivalent of constructing a projective conic from a standard one,  as described in \cite{Cariglia2015}.

In our case the process transforms (\ref{eq:relationship_nonhom}) into
\ben
p_\eta p_v = m a(\eta) p_v^2 + \frac{h^{ij} p_i p_j}{2ma(\eta)} \, .
\een
This dynamical relation can be recovered by studying the null geodesics of the Hamiltonian
\ben
\mathcal{H} = \frac{h^{ij} p_i p_j}{2ma(\eta)} - p_\eta p_v + m a(\eta) p_v^2 \, ,
\een
which is associated with the metric
\ben
ds^2 = m a(\eta) h^{ij} dx_i dx_j - 2ma(\eta) d\eta^2 - 2 d\eta dv \, .
\een
After a translation, $v = \tilde{v} - m \int^\eta a(\eta^\prime) d\eta^\prime$, we obtain
\ben
ds^2 = m a(\eta) h^{ij} dx_i dx_j - d\eta d\tilde{v} \, .
\een
In particular, if the spatial slices are flat then the metric is conformally flat. Its conformal symmetries contain the Schr\"odinger group as a subgroup, providing us with symmetries  also of  the original FLRW Hamiltonian $H_{FLRW}$ in (\ref{FLRWHam}).

Another consequence is that the null geodesics of the Eisenhart-Duval lift just constructed are the same as the null geodesics of the Ermakov-Milne-Pinney spacetime, described in sec.\ref{EMPEDliftsec}.

We conclude this subsection by generalizing our findings to $N$ particles in a FLRW spacetime, subject to their mutual gravitational interaction. The Hamiltonian we use is
\ben
\frac{1}{2a^2} \left[ - p_{\eta}^2 + \sum_{A=1}^N h^{ij} p^A_i p^A_j \right] - \sum_{1\le A < B \le N} \frac{G m_i m_j}{a|\vec{r}^A - \vec{r}^B |}
\, ,
\een
where the $p_i^A$ are the spatial momenta of the individual particles and $\vec{r}^A$ their coordinates.

 Our Hamiltonian relies on the assumption that the proper times of all particles can be synchronized and described by the same parameter. This is possible since the spacetime is spatially homogeneous. As before, we set the Hamiltonian equal to a constant value $-{M^2}/{2}$, calculate the low-energy limit and make the relationship homogeneous in the momenta. We obtain the null Hamiltonian
\ben
\mathcal{H} = - p_\eta p_{\tilde{v}} + \frac{1}{2Ma} \sum_{A=1}^N h^{ij} p^A_i p^A_j - \frac{1}{M} \sum_{1\le A < B \le N} \frac{G m_i m_j}{|\vec{r}^A - \vec{r}^B |} p_{\tilde{v}}^2 \, ,
\een
where again we have redefined the $v$ variable in order to remove the term $Ma(\eta) p_v^2$. Going back to cosmic time, with $p_\eta = a p_t$, we obtain
\ben
\mathcal{H} = a(t) \left[ - p_t p_{\tilde{v}} + \frac{1}{2Ma^2} \sum_{A=1}^N h^{ij} p^A_i p^A_j - \frac{1}{Ma} \sum_{1\le A < B \le N} \frac{G m_i m_j}{|\vec{r}^A - \vec{r}^B |} p_{\tilde{v}}^2 \right] \, ,
\een
which belongs to the same conformal class as the Eisenhart-Duval lift of the Hamiltonian presented in \cite{GibbonsEllis2015}.

This gives rise to the Dmitriev-Zel'dovich equations \cite{DimZel,GP,GibbonsEllis2015}.
If we consider, as is done in almost all numerical
studies  of large scale structure, the motion of $N$  non-uniformities
about a FLRW background metric, they
are governed by the Dmitriev-Zeldovich equations \cite{DimZel,GibbonsEllis2015}
\begin{equation}
  \frac{d}{dt} \Bigl ( a^2(t) \dot {\bf x}_a      \Bigr) = \frac{1}{a(t)}
  \sum _{b\ne a} \frac{Gm_b( {\bf x}_b-{\bf x}_a) } { | {\bf x}_b -{\bf x}_a |} \label{DZ}
  \end{equation}
where $a,b=1,2,\dots,N$ label the non-uniformities and $m_a$ are their masses.
It is simple to verify that if one replaces every ${\bf x}_a$ in (\ref{DZ}) by
\begin{equation}
{\bf x}_a +  {\bf a} + {\bf b} \int _0^t \frac{dt ^\prime}{a(t^\prime) }
\end{equation}
then the Dmitriev-Zeldovich equations are left invariant.

The equations (\ref{DZ}) may be derived from a Lagrangian \cite{Peebles,GibbonsEllis2015}
\begin{equation}
 L= \frac{1}{2} \sum_ {1\le a \le N} a^2(t) m_a {\dot {\bf x}_a}^2 +
\frac{1}{a(t)} \sum_ {1\le a < b \le N} \frac{Gm_am_b}{|{\bf x}_a-{\bf x}_b|}.
\end{equation}

This system Eisenhart-Duval lifts to
\begin{equation}
ds ^2 = \sum_ {1\le a \le N} a^2(t) m_a  d {\bf x}_a^2 + 2 dt dv + 2  \frac{1}{a(t)} \sum_ {1\le a < b \le N} \frac{Gm_am_b}{|{\bf x}_a-{\bf x}_b|} dt^2 \,.
\end{equation}

Up to signs and factors this coincides with the metric (\ref{EM}) as long as $\gamma=0$ and $G=0$.
In other words a free non-relativistic particle in a  background Friedmann universe
lifts to a null geodesic of (\ref{EM}) provided $\gamma =0$. In fact the time-dependent harmonic oscillator in (\ref{EM}) looks very much like a cosmological constant term.

\subsection{The Eisenhart--Duval lift of the Friedmann equations}\label{EisenFriedSec}

The action density of Einstein gravity coupled to a scalar field is
\ben
{\cal L}= \Bigl \{ \frac{1}{16 \pi G}    R -\half g^{\mu \nu} \p_\nu \phi \p_\nu \phi - V(\phi) \Bigr \} \,.
\een
Substituting in the FLWR metric and assuming $\phi$ depends on $t$ we
find that \emph{up to a total time derivative}
\ben
{\cal L}= \frac{6}{1 6 \pi G} (-a \dot a ^2 + K a ) + a^3  (\half \dot \phi ^2 + V(\phi))\,.
\een
Varying ${\cal L}$  with respect to $\phi$ and $a$  yields the field equation for $\phi$
and the following second order ODE  for $a$
\ben
\frac{\ddot a} {a} + \frac{\dot a ^2}{a^2} = -\frac{4 \pi G}{3}
(\dot \phi ^2 - 4 V(\phi) ) \,.
\label{eqn}
\een
One may check that (\ref{eqn}) is a linear combination of the second order Raychaudhuri equation and
the first order Friedmann equation.
If one calculates the Hamiltonian density $\cal H$ of $\cal L$ one
finds that the two canonical momenta are
\ben
p_\phi= a^3 \dot \phi \,,  \qquad p_a =- \frac{3}{4 \pi G} a \dot a
\,,\een
and considered  as a  function of the velocities $(\dot a, \dot \phi)$,
we find that
\ben
{\cal H}= \frac{6}{1 6 \pi G} (-a \dot a ^2 - K a) + a^3 ( \half \phi ^2 + V(\phi) )\,.
\een
One may check that the Friedmann ODE is the same as the constraint
\ben {\cal H}=0\,.\label{HamCon}
\een
To understand better what is happening we note that in setting $g_{tt}=-1$
in the Friedmann metric we have used up some gauge freedom and as a consequence the $\cal L$ does not depend upon sufficiently many independent variables in order to obtain
the full set of equations of motion. To rectify this problem, on considers the more general metric
\ben
ds^2 =-N^2(t) dt ^2 + a^2(t) d \Sigma_K ^2
\een
which however differs from our previous form
by no more than a \emph{temporal diffeomorphism}
or \emph{clock regraduation}.The function $N(t)$ is called the {\it lapse function}. It measures the lapse of time relative to cosmic time. Now
\ben
\tilde {\cal L}= \frac{6}{1 6 \pi G} (- N^{-1}a \dot a ^2 +N  K a ) +
a^3 ( N^{-1} \half \dot \phi ^2 + N V(\phi)) \,.\label{gauge}
\een
We see that the lapse function $N$ enters (\ref{gauge}) as a
\emph{Lagrange multiplier}
and  varying  the Lagrangian $\tilde {\cal L}$
with respect to $N$ and then setting
$N=1$ yields the Friedmann equation, i.e. the
Hamiltonian constraint (\ref{HamCon}).
In fact in terms of the canonical  momenta  we have
\footnote{From now on we drop the tildes.}
\ben
{\cal H}= N \Bigl \{  - \frac{2 \pi G}{3} \frac{p_a^2}{a} - \frac{3 K}{8 \pi G}
+ \half \frac{p_\phi ^2 }{a^3}  + a^3 V(\phi) \bigr \} \,. \label{HaNiltonian}
\een
We are now in the situation to perform an Eisenhart-Duval lift
or Marsden-Weinstein oxidation, we set
\ben
p_t= -{\cal H}
\een
and  introduce an ignorable momentum $p_v$ and rewrite  (\ref{HaNiltonian}) as
\ben
2 H_{\rm{Eisenhart-Duval}}=  2p_t p_v +  2 N \Bigl\{ - \frac{2 \pi G}{3} \frac{p_a^2}{a} - \frac{3 K}{8 \pi G} p_v^2
+ \half \frac{p_\phi ^2 }{a^3}  + a^3 V(\phi) p_v^2 \Bigr\}=0 \,.
\een

Note that if we set $N=1$ and compute the 4- metric then
the coefficients of $da ^2 $ and $ d \phi ^2$ have the opposite sign.
If we regard $N$ as a further coordinate, then
we have a Galilean or Newton-Cartan metric.

\vspace{0.5cm}

\section{Conclusion}\label{concl}

Let us recapitulate the results obtained in this work:

\begin{itemize}
\item
Any Newtonian mechanical system can be described in terms of the Eisenhart-Duval metric which  \textit{solves} the Einstein equations (\ref{EE1}). The key ingredient involved in the construction is the energy--momentum tensor (\ref{EMT}) built out of the covariantly constant null Killing vector field (\ref{Xi}) and a proper energy density function.
\item
The celebrated Ermakov-Lewis invariant of a time-dependent harmonic oscillator
can be obtained in purely geometric way by applying the Eisenhart-Duval lift.
\item
A cosmological extension of the Eisenhart-Duval metric is constructed by properly incorporating into the scheme a cosmic scale factor and the energy-momentum tensor. The evolution of spacetime is governed by the Ermakov-Milne-Pinney equation.
\item
Killing isometries include spatial translations and rotations, Newton--Hooke boosts and translation in the null direction.
\item
The algebra of vector fields which involve both Killing and conformal Killing vectors is identified with $so(2,2+d)$, the conformal Newton-Hooke algebra being its subalgebra.
\item
Geodesic motion in Ermakov-Milne-Pinney cosmoi is described.
\item
The Eisenhart-Duval lift of geodesics in the Friedmann-Lema\^\i tre-Robertson-Walker spacetimes is found and then generalized to the Dmitriev-Zel'dovich equations.
\item
The derivation of the Friedmann equations within the framework of the Eisenhart-Duval lift is presented.
\end{itemize}

\begin{acknowledgments}
  We would like to thank an anonymous referee for informing us that Ermakov's paper  was preceeded by one of  Adolph Steen \cite{Steen,Redheffer} and therefore it would be more appropriate to speak of the Steen-Ermakov-Milne-Pinney equation. MC was funded by the CNPq under project 303923/2015-6, and by a \textit{Pesquisador Mineiro} project n. PPM-00630-17. AG was supported by the Tomsk Polytechnic University competitiveness enhancement program.
\end{acknowledgments}


\end{document}